\documentclass[9pt,aps,prd,nofootinbib,supercriptaddress]{revtex4}
\usepackage{amsmath}
\usepackage{amsfonts}
\usepackage{graphicx}
\usepackage[T1]{fontenc}
\usepackage[latin9]{inputenc}
\usepackage{amssymb}
\usepackage{mathrsfs}

\def\be{\begin{equation}}
\def\ee{\end{equation}}
\def\bea{\begin{eqnarray}}
\def\eea{\end{eqnarray}}

\def\cal H{\,{\mathcal  H}}

\newcommand{\sdelta}[1]{\!\delta^{\,3}(\mathbf{#1})}

\def\picube{(2\pi)^3}

\def\cmm2{{\,\rm cm^{-2}}}
\def\cm2{{\,{\rm cm}^2}}
\def\cmm3{{\,{\rm cm}^{-3}}}
\def\gcmm3{{\,{\rm g\,cm^{-3}}}}

\def\bkone{\mathbf k_1}
\def\bktwo{\mathbf k_2}
\def\bkthree{\mathbf k_3}
\def\bkfour{\mathbf k_4}

\def\picube{(2\pi)^3}

\begin{document}

\title{A Diagrammatic Approach to Scalar Field Correlators during Inflation}

\author{G. Petri$^{1}$}

\affiliation{$^1$  Department of Civil and Environmental Engineering and Institute of Mathematical Sciences, \\
Imperial College, London SW7 2AZ, UK\\
}

\begin{abstract}

\noindent We consider a self-interacting scalar field in a de Sitter background
and deal with the associated infrared divergences in a purely diagrammatic way using the in-in formalism.
In the particular case of a large N  O(N) invariant scalar field theory with quartic self-interactions
we recover the result that the connected four-point correlation function, which is a signal of non-Gaussianity,
is non-perturbatively enhanced with respect to its tree-level value.
\end{abstract}

\maketitle

\section{Introduction}

\noindent
The detection of non-Gaussianity (NG) \cite{review} in  the   cosmological perturbations generated during inflation \cite{lrreview}
 has become one of the  primary targets of forthcoming experiments measuring the properties of the Cosmic Microwave
Background (CMB) and matter fluctuations. NG is originated by  the self-interactions of  the fields involved in the
inflationary dynamics plus those induced by gravity. The effect of loop corrections on  the physical observables
generated during a de Sitter epoch of exactly exponential expansion
have attracted much attention
in the past \cite{hu1,hu2,sasaki1,sasaki2,starob,Mukhanov:1996ak,Abramo:1997hu}  and more recently
\cite{boy1,boy2,olandesi,
Sloth:2006az,Sloth:2006nu,Weinberg:2005vy,wein2,seery1,seery2,lythbox,bmprs,slothriotto,kari}, especially as far as the
resummation  of infra-red (IR) divergences is concerned.   IR divergences appear because of the
cumulative effects of the superhorizon perturbations which are continuously generated during the
de Sitter stage.
Different approaches have been put forward to deal with them. The 2PI (Two-Particle-Irreducible)  formalism has been adopted
in Refs. \cite{hu1,hu2}, while  a stochastic field theory method has originally been used in Ref. \cite{starob} where
the underlying idea is that the IR part of a scalar field may be considered as  a classical space-dependent stochastic
field satisfying a local Langevin equation. The stochastic noise terms arise from the quantum
fluctuations which becomes classical at horizon crossing and then contribute to the background.
An hybrid method, combining the stochastic approach and the out-of-equilibrium field theory techniques of the in-in formalism
\cite{schwinger1961} to solve the gap equation descriving the time-dependent evolution of the two-point correlator,
has been recently used in Ref. \cite{slothriotto} for  a self-interacting $O(N)$ model in the limit of large $N$ and it was
shown that the connected four-point correlator, the so-called trispectrum,
is non-perturbatively enhanced with respect to its tree-level value.

In this paper we adopt a purely diagrammatic approach based on the in-in formalism
to analyze the   IR divergences. We restrict ourselves to a
quartic self-interacting scalar field in a de Sitter background and show that the IR resummation of a certain class of diagrams
occurs. Generalizing the computation to  $N$ scalar fields subject to an  $O(N)$ symmetry with large $N$, we recover the
results of Ref. \cite{slothriotto}.  In this sense, our results
should be considered complementary to alternative approaches, {\it e.g.} the stochatic approach.

The paper is organized as follows.
In Section \ref{CTP} we summarize the in-in formalism and the  Feynmann rules needed to calculate the higher order
corrections for the scalar field correlators. In Section \ref{mieiconti} we explicitly perform the loop calculations.
In Section IV, we discuss the particular case of $O(N)$ symmetric model and the trispectrum. In Section \ref{conclusions}
 we summarize and conclude our work.


\section{Self-interacting Scalar Field in de Sitter background}\label{CTP}
\subsection{Closed Time Path Formalism}
\noindent
We use the in-in formalism, also dubbed Closed Time Path (CTP) formalism,  to calculate the correlation functions of a scalar field in a de Sitter background. Following Schwinger \cite{schwinger1961,calzetta}, we introduce the two external sources $J^+(x)$ and $J^-(x)$ and consider the quantity
\be\label{Zjj}
Z[J^+, J^-] = {}_{J^-} \langle0_-|0_+\rangle {}_{J^+} \, .
\ee
The vacuum evolves indipendently under two sources $J^+$ and $J^-$. We may rewrite the latter as
\begin{eqnarray}\label{Zjj2}
Z[J^+, J^-] &=& \int D\phi \left\langle 0_- \left| \tilde{T}{\rm exp} \left[-i \int_{-\infty}^{t*} dt \int d^3x  J^-(x)\phi_H(x)\right]\right|\phi\right\rangle \nonumber \\
&&  \left\langle \phi \left| T {\rm exp} \left[i \int_{-\infty}^{t*} dt \int d^3x  J^+(x)\phi_H(x)\right]\right|0_- \right\rangle \, ,
\end{eqnarray}
where $\tilde{T}$ denotes antitemporal order. Here $|\phi\rangle$ is an element of a complete, orthonormal set of common eigenvectors of the field operators at some late time $t^*$,
\be
\phi_H({\bf x},t)|\psi\rangle = \Phi({\bf x})|\phi\rangle.
\ee
From the definitions (\ref{Zjj}) and (\ref{Zjj2}), one can obtain the following relations \cite{calzetta}:
\be
Z[J,J] = 1, \quad Z[J^+, J^-] =\quad (Z[J^-,J^+])^* \, ,
\ee
and
\begin{eqnarray}
(-i)^{n-m}  \frac{\partial^{n+m} Z[J^+, J^-]}{\partial J^-(x_1)\ldots \partial J^-(x_m) \partial J^+ (y_1)\ldots \partial J^+ (y_n)} \Bigg|_{J^+, J^- = 0}  =  \nonumber \\
\langle0_-|\tilde{T}[\phi_H(x_1)\dots \phi_H(x_m)] T[\phi_H(y_1)\dots \phi_H(y_n)]|0_-\rangle  .
\end{eqnarray}
The expectations value can be be obtained by variation of the sources $J^+$ and $J^-$. In particular for a time-dependent Hamiltonian system $H(t)$ that starts in a state $|in\rangle$ at time $t_{in}$, we can write the expectation value as:\\
\begin{equation}
\langle Q(t) \rangle = \left\langle {\rm in}\left| \left[\bar{\rm T}
  \exp\left( i \int_{t_{\rm in}}^{t} dt' \, H(t') \right) \right] Q
  \left[{\rm T} \exp \left( -i \int_{t_{\rm in}}^t dt'\, H(t') \right)
  \right] \right| {\rm in} \right\rangle \, .
\end{equation}
\\
Now we move to a curved space, namely to a de Sitter background. We write the Lagrangian density for a scalar field with potential $V(\phi)$ as
\be
\mathscr{L}[\phi] = \sqrt{-g} \left( g^{\mu\nu} \frac{1}{2} \partial_{\mu}\phi\partial_{\nu}\phi -\frac{1}{2}m^2\phi^2 -\frac{1}{2}\xi R\phi^2 - V(\phi)\right) +\delta \mathscr{L} \, ,
\ee
where the metric has signature $-+++$, $\xi$ is the conformal parameter and $\delta\mathscr{L}$ is the counterterm. Choosing $m=0$ and $\xi = 0$ we select a massless minimally coupled scalar field.
The generating functional becomes \cite{olandesi}\\
\begin{eqnarray}
Z[J_+,J_-,\rho(t_{\rm in})]& = &  \int \mathcal{D} \phi^+_{\rm in} \mathcal{D}
\phi^-_{\rm in} \langle \phi^+_{\rm in} | \rho(t_{\rm in}) |
\phi^-_{\rm in} \rangle \; \\
& &  \int_{\phi^+_{\rm in}}^{\phi^-_{\rm in}} \mathcal{D} \phi^+
\mathcal{D} \phi^- e^{ i \int_{t_{\rm in}}^t dt' \int d^3 x
\left(\mathcal{L}[\phi^+] - \mathcal{L}[\phi^-] + J_+ \phi^+ + J_-
\phi^- \right)} \nonumber\, . \label{ctpz}
\end{eqnarray}
The path integral on the second line can be written in short-hand notation as
\begin{equation}
\int \mathcal{D} \phi \; \exp\left[ i \int_{\mathcal{C}} dt' \int d^3 x \,
  \left( \mathcal{L}[\phi] + J \phi \right) \right],
\end{equation}
where $\mathcal{C}$ is the so-called Schwinger-Keldysh contour which
runs from $t_{\rm in}$ to $t$ and back. The field $\phi$ and source
$J$ are split up in $\phi^+$, $J_+$ on the first part of this contour,
and $\phi^-$, $J_-$ on the second part, with the condition $\phi^+(t)
= \phi^-(t)$.
\begin{figure}
\centering
\includegraphics[width=.4\textwidth]{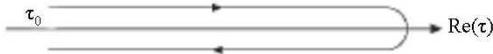}
\caption{An example of Keldysh Contour $\mathcal{C}$}
\end{figure}
\\
To calculate perturbatively correlation functions we need to have the free two-point functions. There are four possible time orderings and, using eq.~(\ref{ctpz}) one obtains:\\
\begin{eqnarray}
G^{-+}(x,y) &=  i \langle \phi(x) \phi(y) \rangle^{(0)},
\label{gmp} \\
G^{+-}(x,y) &=  i \langle \phi(y) \phi(x) \rangle^{(0)},
\end{eqnarray}
and
\begin{eqnarray}
G^{++}(x,y) &=&  i \langle {\rm T} \phi(x) \phi(y) \rangle^{(0)}  \\
& = & \theta(x_0-y_0) G^{-+}(x,y) + \theta(y_0-x_0) G^{+-}(x,y),\nonumber \\
G^{--}(x,y) &=&  i \langle {\rm \bar{T}} \phi(x) \phi(y) \rangle^{(0)} \\
&= & \theta(x_0-y_0)
G^{+-}(x,y) + \theta(y_0-x_0) G^{-+}(x,y),\nonumber
\end{eqnarray}
where by the superscript $(0)$ we denote the free field correlation functions. They obey the important identity
\begin{equation}
G^{++}(x,y) + G^{--}(x,y) = G^{-+}(x,y) + G^{+-}(x,y), \label{gid}
\end{equation}
and they can be put together in a matrix:
\begin{equation}
\mathbf{G}(x,y) = \left( \begin{array}{cc}
G^{++}(x,y) & G^{+-}(x,y) \\
G^{-+}(x,y) & G^{--}(x,y) \end{array} \right).
\end{equation}
Note that the two point functions depend on the initial conditions via the dependence on $\rho(t_i)$ of the generating functional eq.~(\ref{ctpz}). It is useful to transform the $\phi^+$ and $\phi^-$ fields to a different basis, which
is a variation of the Keldysh basis:\\
\begin{equation}\label{fieldchange}
\left( \begin{array}{c} \phi_C \\ \phi_\Delta \end{array}\right)  =
\left( \begin{array}{c} (\phi^+ + \phi^-)/2 \\ \phi^+ - \phi^-
\end{array} \right) =
\mathbf{R} \left( \begin{array}{c} \phi^+ \\ \phi^-
\end{array} \right), \ee
with
\be
\mathbf{R} = \left(
\begin{array}{cc} 1/2 & 1/2 \\ 1 & -1 \end{array} \right) .
\end{equation}\\
The free two point functions in this basis can easily be obtained by
the transformation
\begin{equation}\label{matricetta}
\mathbf{G}_K(x,y) = \mathbf{R} \mathbf{G}(x,y) \mathbf{R}^T = \left(
\begin{array}{cc} i G_C (x,y) & G_R (x,y) \\ G_A(x,y) & 0 \end{array}
\right),
\end{equation}
The "$G_{\Delta\Delta}$" propagator in the matrix (\ref{matricetta}) (the element (2,2) of $\mathbf{G}_K(x,y)$) is identically zero due to eq.~(\ref{gid}), as can be seen by performing directly the product. Finally the $G_R$ and $G_A$ two point functions are
called the retarded and advanced propagators and $G_A(x,y) = G_R(y,x)$.

\subsection{Feynmann Rules for $\phi^4$}
We choose the potential to be $V(\phi)=\frac{\lambda}{4!}\phi^4$. The Lagrangian density becomes:
\begin{displaymath}
\mathscr{L}[\phi] = \sqrt{-g} \left( g^{\mu\nu} \frac{1}{2} \partial_{\mu}\phi\partial_{\nu}\phi  + \frac{\lambda}{4!}\phi^4\right)
\end{displaymath}
We perform now the field transformation as in eq.(\ref{fieldchange}), obtaining
\begin{equation}
\mathscr{L}[\phi_C, \phi_\Delta] = \sqrt{-g}\left[g^{\mu\nu}\partial_\mu\phi_C\partial_\nu\phi_\Delta - \frac{\lambda}{4!}\left(4\phi^3_C\phi_\Delta 
+ \phi_C\phi_\Delta^3\right)\right]
\end{equation}
We notice that the theory has two vertices. From now on we will utilize the conformal time $\tau$, defined as $\tau = -\int_t^{\infty} dt'/a(t')$. Note that, as a function of $\tau$, the scale factor is {$a(\tau)=-(H\tau)^{-1}$}. The free two-point functions in the late time limit are \cite{olandesi}:
\begin{eqnarray}
\mathrm{G}_C^{(0)}(k,\tau_1,\tau_2) &=& \frac{H^2}{2 k^3}, \\
G_R(k,\tau_1,\tau_2) &= & \theta({\tau_1}-{\tau_2}) \frac{H^2}{3}({\tau_1}^3-{\tau_2}^3)  \label{freegr}
\end{eqnarray}
and $\mathrm{G}_A^{(0)}(k,\tau_1,\tau_2) = \mathrm{G}_R^{(0)}(k,\tau_2,\tau_1)$. The two point functions depend only on the length of the spatial momentum $k =|\mathbf{k}|$. \\
Following \cite{olandesi},  we represent the $\phi_C$ field with a full line and the $\phi_\Delta$ field with a dashed line and so we can write the Feynman rules for the two-point functions as\\ 

\begin{figure}[!h]
\centering
\includegraphics[width=.15\textwidth]{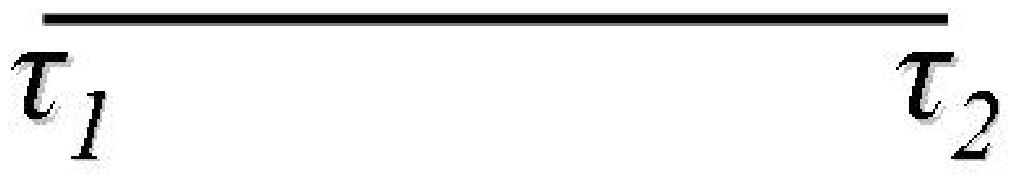} \\ $\mathrm{G}_C^{(0)}(k, \tau_1, \tau_2)$ \, , \end{figure}

\begin{figure}[!h] \centering \includegraphics[width=.15\textwidth]{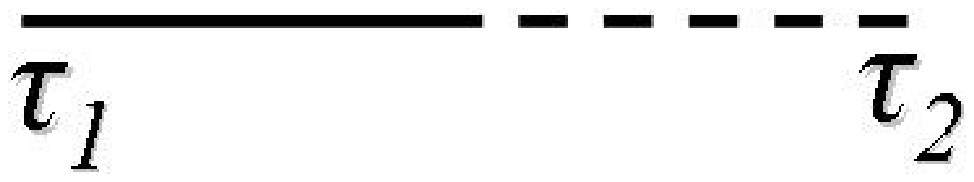}  \\ $-i \mathrm{G}_R^{(0)}(k,\tau_1,\tau_2) = -i
\mathrm{G}_A^{(0)}(k,\tau_2,\tau_1)$\, .
\end{figure}

We have two different vertices. One contains three powers of $\phi_C$ 
and one of $\phi_\Delta$ so we draw it with three full lines and one dashed line. The other instead contains one power of $\phi_C$ and three of $\phi_\Delta$, hence a vertex with three dashed lines and one full line. Since we are in a de Sitter background, $\sqrt{-g}=a^4(\tau)$ and the vertices become:

\begin{figure}[!h]
\centering
\includegraphics[height=2 cm]{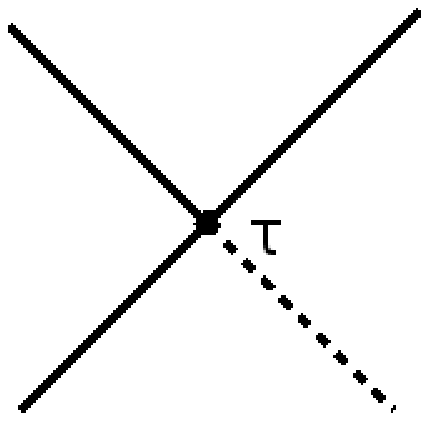} \hspace{4cm}  \includegraphics[height=2cm]{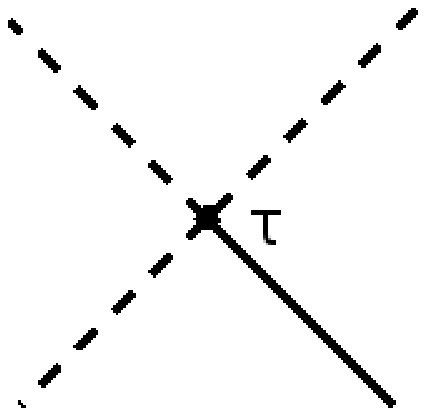}\end{figure}

\hspace{4cm} $-i a^4(\tau) \lambda{\phi_C}^3\phi_\Delta$ \hspace{3.5cm}  
$-i a^4(\tau) \frac{\lambda}{4}\phi_C \phi^3_\Delta $\\
When a two point function is attached to a vertex, the corresponding
time has to be integrated over, so we get a $\int d\tau$, while for a  closed loop we get an
integral over the internal spatial momentum $\int d^3p/(2 \pi)^3$. \\
\begin{figure}[!ht]
\centering
\includegraphics[width=0.3\textwidth]{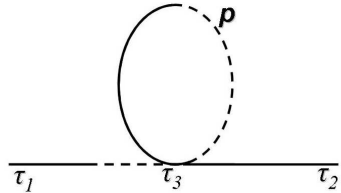}\label{Grloop}
\caption{The loop is formed by a retarded propagator $G_R$ starting and ending at time $\tau_3$. It is identically zero.}
\end{figure}
Considering the form of $G_R$ we can already exclude the presence of loop with mixed lines, like in figure 2. Indeed,
such a loop would close a retarded propagator $G_R$ on the same time $\tau = \tau_1 = \tau_2$ but, due to the embedded $\theta(\tau_1 -\tau_2)$, it vanishes. So the only possible loop that we can build with our set of Feynmann's rules is made of a full line. It is given by
\be\label{loop}
\Lambda(\tau) = \int \frac{d^3p}{(2\pi)^3} \mathrm{G}_C^{(0)}(p, \tau, \tau)\, ,
\ee
where again by the superscript $(0)$ we mean the free correlation functions.
Since we will calculate correlation functions at higher orders, the superscript $(i)$ will help us to
keep track of which order are we considering at each moment.
As argued in \cite{calzetta}, primitively divergent graphs contain only vertices of the same type. If there were vertices of different type, then at least two internal lines would be retarded propagators, the corresponding momenta would be on shell, the corresponding loop integral would be finite and the graph would not have been primitively divergent. Now the graphs of the in-in effective action with all vertices of the same sign are just the graphs of the in-out theory plus their complex conjugates, so the primitive divergences must be the same. Once the primitive divergences are controlled, it is only a matter of combinatorics to show that the overlapping divergences disappear as well.

\section{Higher Order Correlation Functions for $\phi$} \label{mieiconti}
\noindent
Since we are interested in the the IR modes, for which  ($-k\tau\ll 1$),
the free two-point functions $G_C(k,\tau_1,\tau_2)$ and $G_R(k,\tau_1,\tau_2)$ can be expanded in powers of $k\tau$ \cite{olandesi}:
\begin{equation}\label{Gc0}
\mathrm{G}_C^{(0)}(k,\tau_1,\tau_2)= \frac{H^2}{2k^3},
\end{equation}
\begin{equation}\label{Gr0}
\mathrm{G}_R^{(0)}(k,\tau_1,\tau_2)= \theta(\tau_1 - \tau_2)\frac{H^2}{3k^3}
[k^3(\tau_1^3 - \tau_2^3)].
\end{equation}
In the in-in formalism there are two vertices but we focus only on the one with three full lines and one dashed line. The reason is that the $\langle\phi_C\phi_C\rangle = G_C$ has a momentum dependence $k^{-3}$ and thus is divergent in the infrared, while the $\langle \phi_C \phi_\Delta\rangle = G_R$ does not. Moreover, we note that for a vertex with three dashed lines it is not possible to build loops since they vanish identically. 

\subsection{First Order Diagrams}\label{onetadGc}
\noindent
The simplest correction to the free propagators $\mathrm{G}_C^{(0)}$ and $\mathrm{G}_R^{(0)}$ comes from the graphs with a single tadpole. The only contribution comes from the graph with the full line loop thus figure \ref{Gc1Loop} is the only first order correction to $\mathrm{G}_C^{(0)}$:
\begin{figure}[!hbtp]
\includegraphics[width=.3\textwidth]{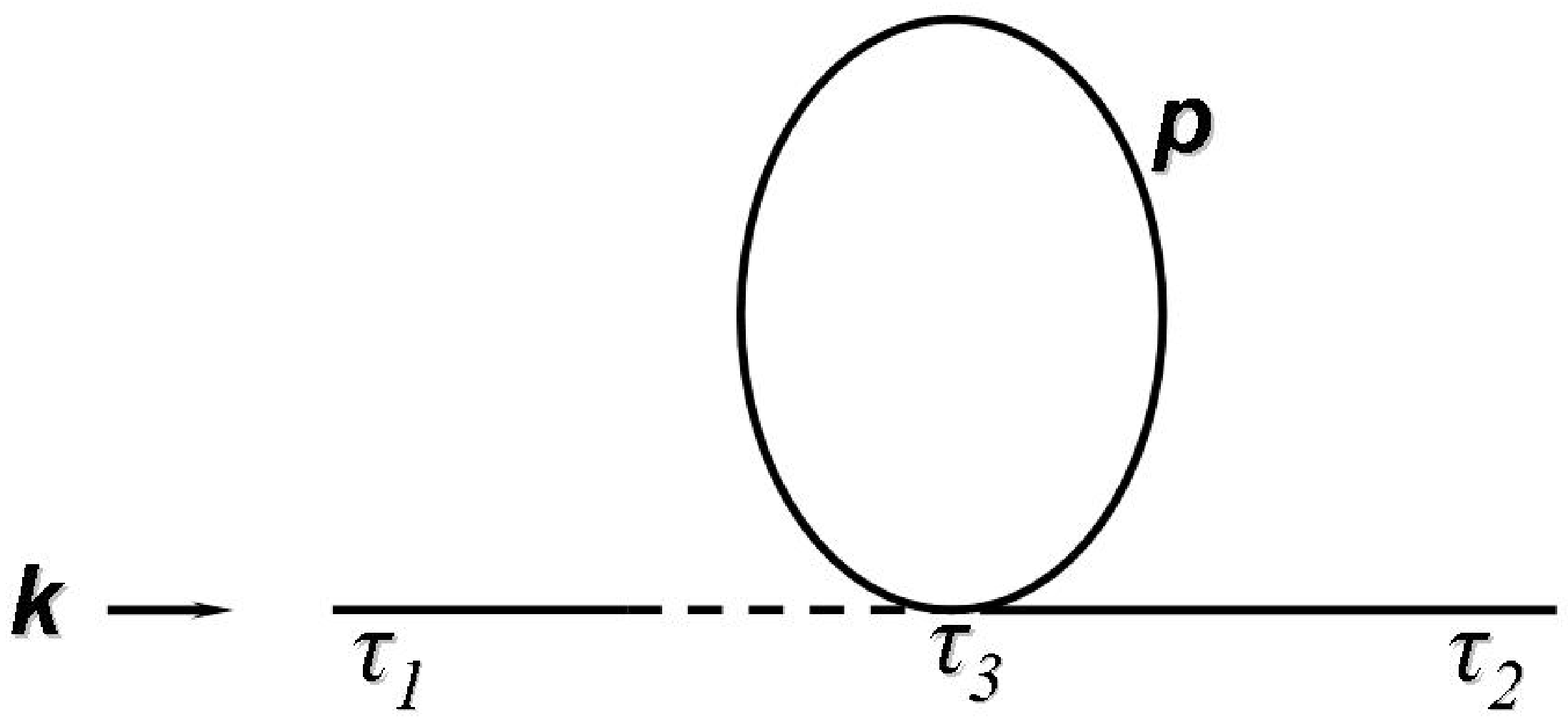}  \includegraphics[width=.3\textwidth]{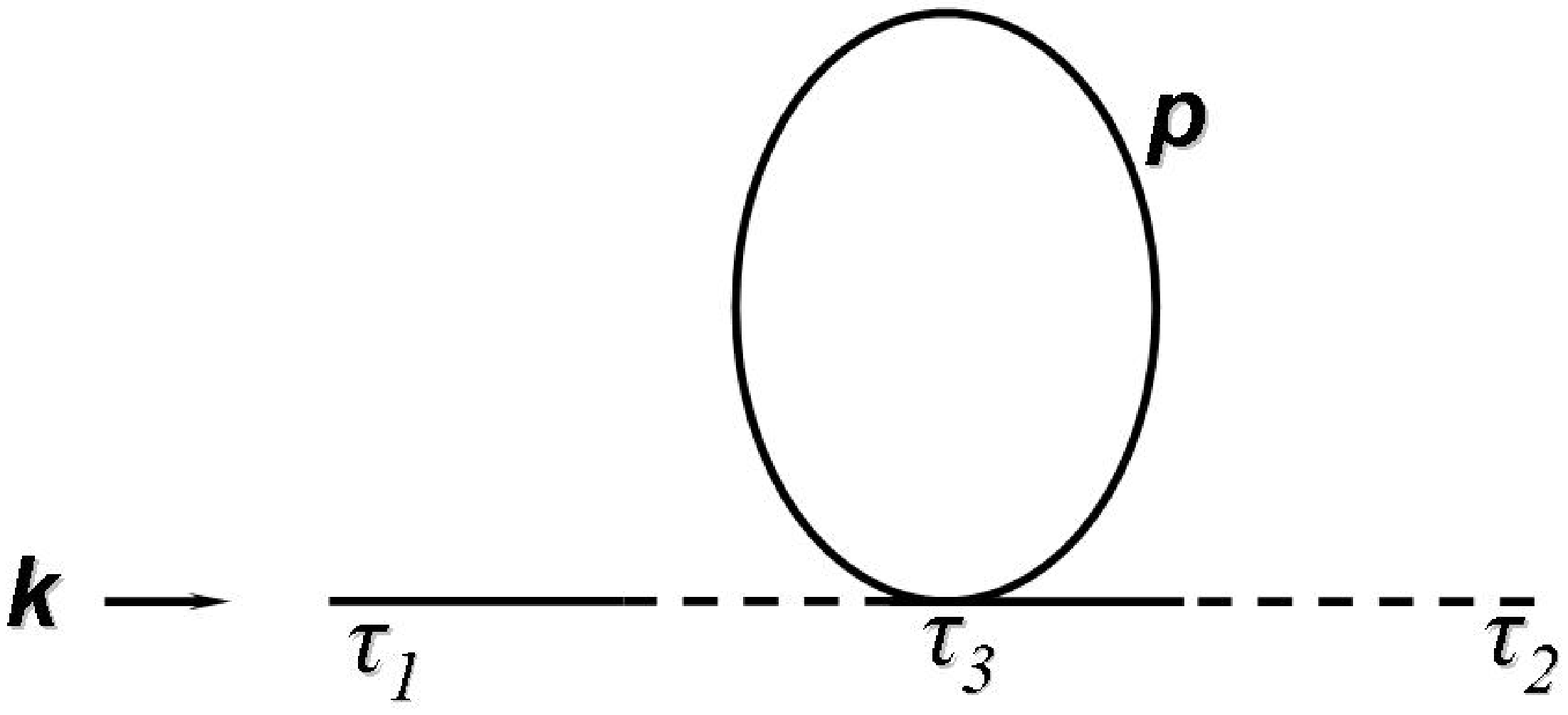}
\caption{One tadpole $G_C$ and $G_R$ propagators}\label{Gc1Loop}

\end{figure}
Due to the simmetry of $\mathrm{G}_C$ we must consider also the mirror diagram and
using the expressions (\ref{Gc0}) and (\ref{Gr0}), we obtain integrals of the form

\begin{equation}\label{Gc1T}
  \int_{-\frac{1}{k}}^{\tau_1} d\tau_3  a^4(\tau_3) \mathrm{G}_R^{(0)}(k,\tau_1,\tau_3)\int \frac{d^3p}{(2\pi)^3} \mathrm{G}_C^{(0)}(p,\tau_3,\tau_3) \mathrm{G}_C^{(0)}(k, \tau_3, \tau_2)  \, ,
\end{equation}
where $p$ is the internal momentum of the tadpole and $k$ the momentum flowing in the diagram.
We set the inferior limit of integration to $-\frac{1}{k}$ instead of $-\infty$, because we are interested in following perturbations from the moment of horizon exit  up to some later time $\tau$. The horizon exit time is given by the
condition $-k\tau_h = 1$ and so $\tau_h = -\frac1k$.
The tadpole integral over $d^3p$ is divergent but can be in general regularized choosing appropriate infrared and ultraviolet cutoffs, $\Lambda_{IR}$ and $\Lambda_{UV}$,
\begin{equation}
\Lambda \equiv \int \frac{d^3p}{(2\pi)^3} \mathrm{G}_C^{(0)}(p,\tau_3,\tau_3) =
\frac{H^2}{(2\pi)^2}  \ln \left(\frac{\Lambda_{UV}}{\Lambda_{IR}}\right)\, .
\end{equation}
In our context the choice of the cut-offs is rather natural.  The IR cut-off $\Lambda_{IR}$ is proportional to
$a_i H$, where $a_i$ is the scale factor at the beginning of inflation and $H$ is the Hubble rate,
while $\Lambda_{UV}$ is equal to $k$, therefore the logarithm is proportional to the total number of $e$-folds from the
beginning of inflation to the time the mode $k$ exits the horizon.
Before performing the calculation we must also consider the coefficient in front of the graph coming from Wick's theorem. We have $\phi_\Delta(\phi_C)^3$ from the vertex  and the external legs, $\phi_C(\tau_1)$ and $\phi_C(\tau_2)$. So there are three possibilities for contracting $\phi(\tau_2)$ with one of the $\phi_C$ of the vertex and one for contracting $\phi_C(\tau_1)$ with the vertex's $\phi_\Delta$, which sum up to 3 in front of the graph.
Performing the calculation and considering also the mirror graph, at leading order we obtain
\begin{eqnarray}\label{gc1}
\mathrm{G}_C^{(1)}(k, \tau_1, \tau_2) &
\simeq & \frac{H^2}{2 k^3} \frac{\lambda \Lambda}{H^2} (\ln(-k\tau_1) + \ln(-k\tau_2)) \, .
\end{eqnarray}
The retarded propagator at one loop has no mirror graph due to the oddness of the $G_R$ under exchange of times.
The Wick contraction factor is again 3. Then, at leading order for a one-loop $G_R$ graph with $\tau_1 > \tau_2$:
\begin{eqnarray}\label{gr1}
-i\mathrm{G}_R^{(1)}(k,\tau_1,\tau_2)  & = & \theta(\tau_1 - \tau_2) \frac{i H^2}{3} \frac{\lambda \Lambda}{ H^2} ({\tau_1}^3 + {\tau_2}^3) \ln\left(\frac{\tau_1}{\tau_2}\right)\, .
\end{eqnarray}
These results coincide with those in Refs.  \cite{Sloth:2006az,Sloth:2006nu} and show the IR divergences due to ther
cumulative effects.
\subsection{Higher Order Corrections}
To try to cure the divergences we need to proceed to higher orders. Already at the second order, three graph topologies can be identified, the tadpole chain (e.g. fig. \ref{G2l}),  the tower graphs(e.g. fig. \ref{tower2l})
and the sunrise (e.g. fig. \ref{sunrise2l}).

\subsubsection{Tadpole Chain Graphs}
\begin{figure}[!hbtp]
\centering
\includegraphics[width=.3\textwidth]{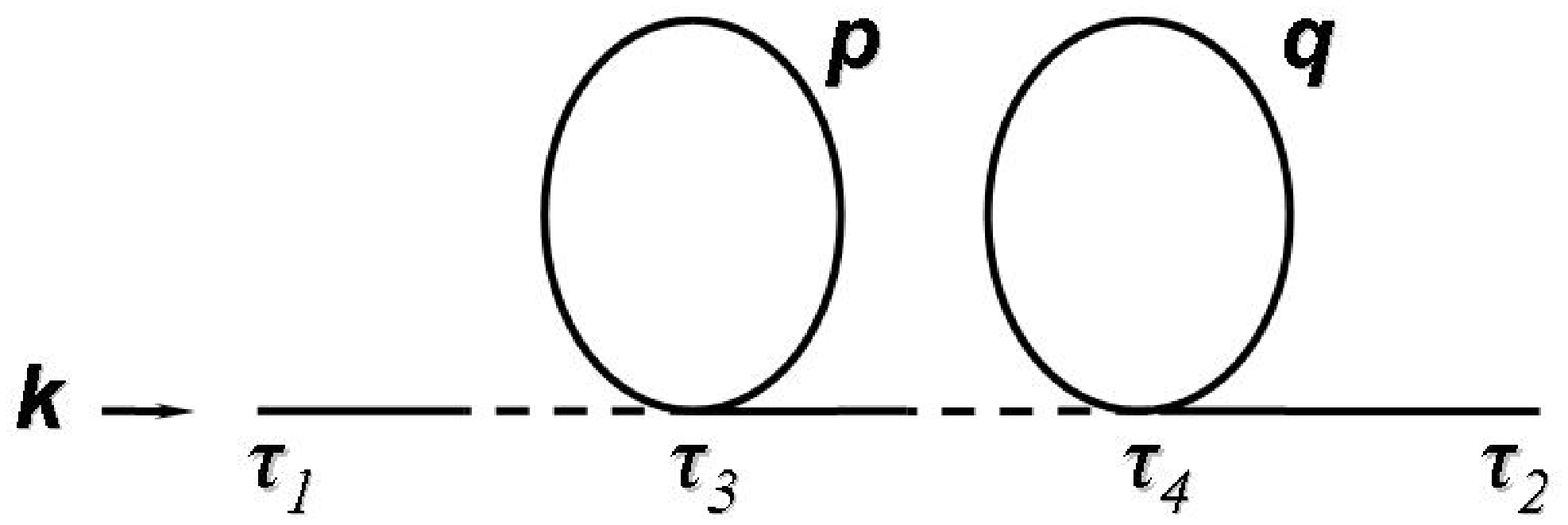} \includegraphics[width=.3\textwidth]{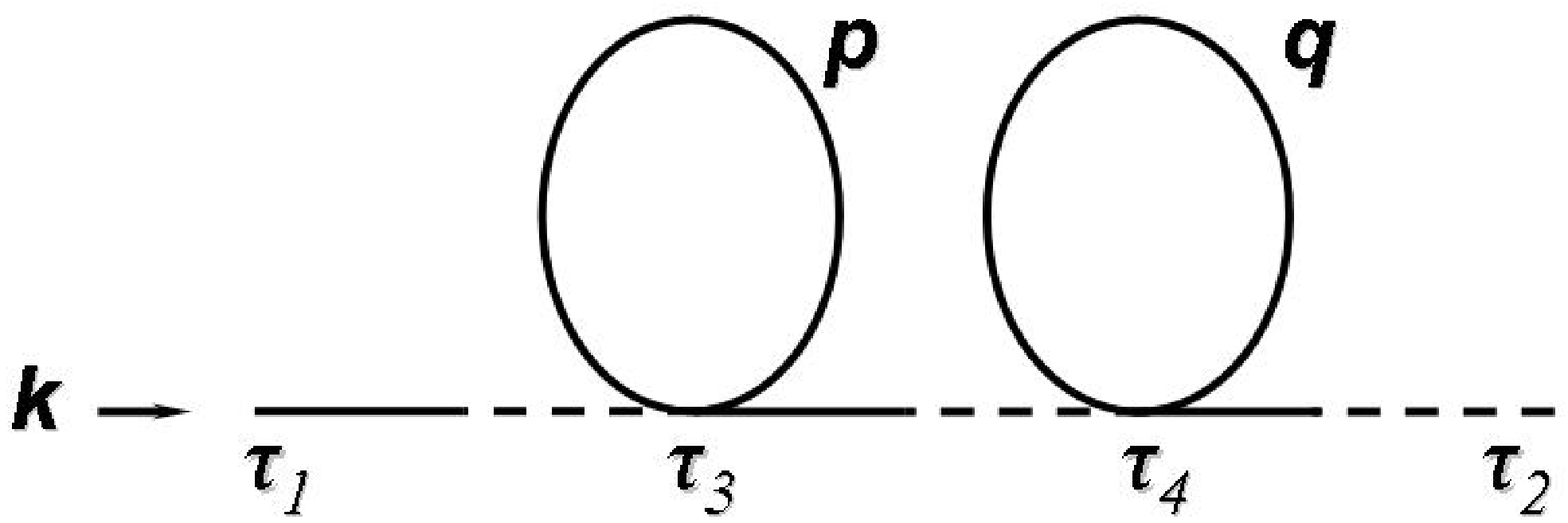}
\caption{Two-tadpole $G_C$ and $G_R$ propagators.}\label{G2l}
\end{figure}

\begin{figure}[!hb]
\centering
\includegraphics[width=.3\textwidth]{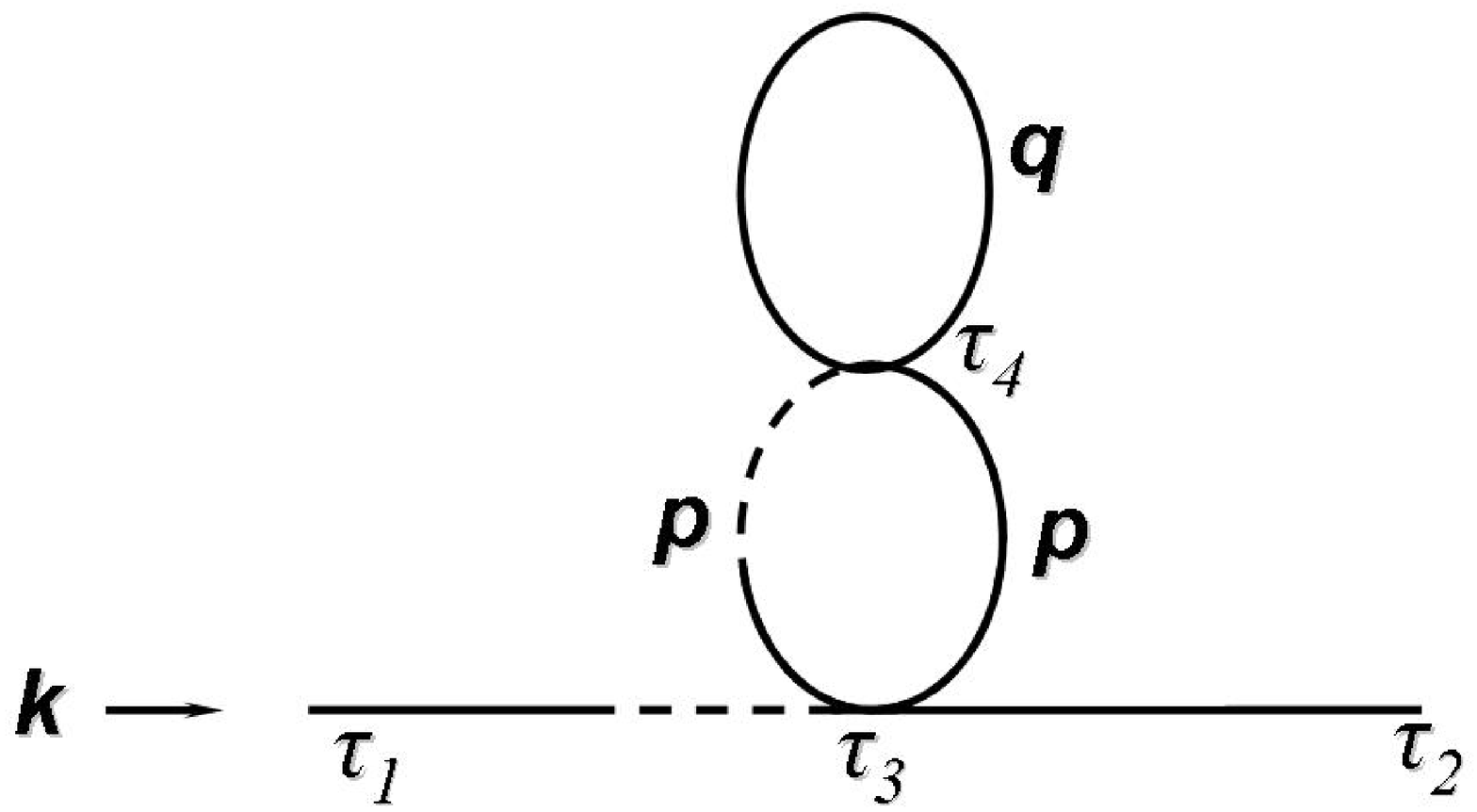} \includegraphics[width=.3\textwidth]{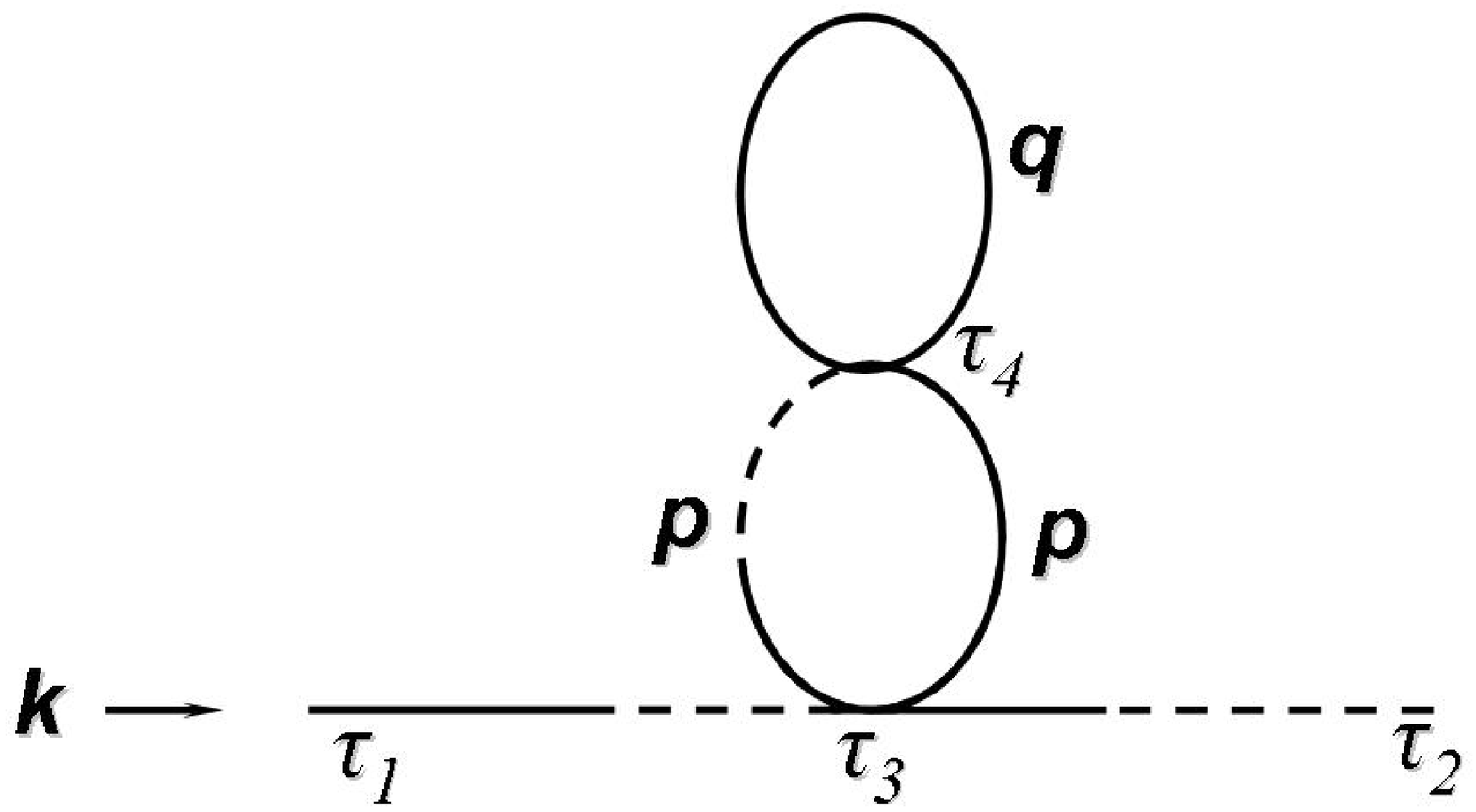}
\caption{Second order tower diagrams  for $G_C$({\it left}) and $G_R$({\it right}).}\label{tower2l}
\end{figure}

\begin{figure}[!h]
\includegraphics[width=.3\textwidth]{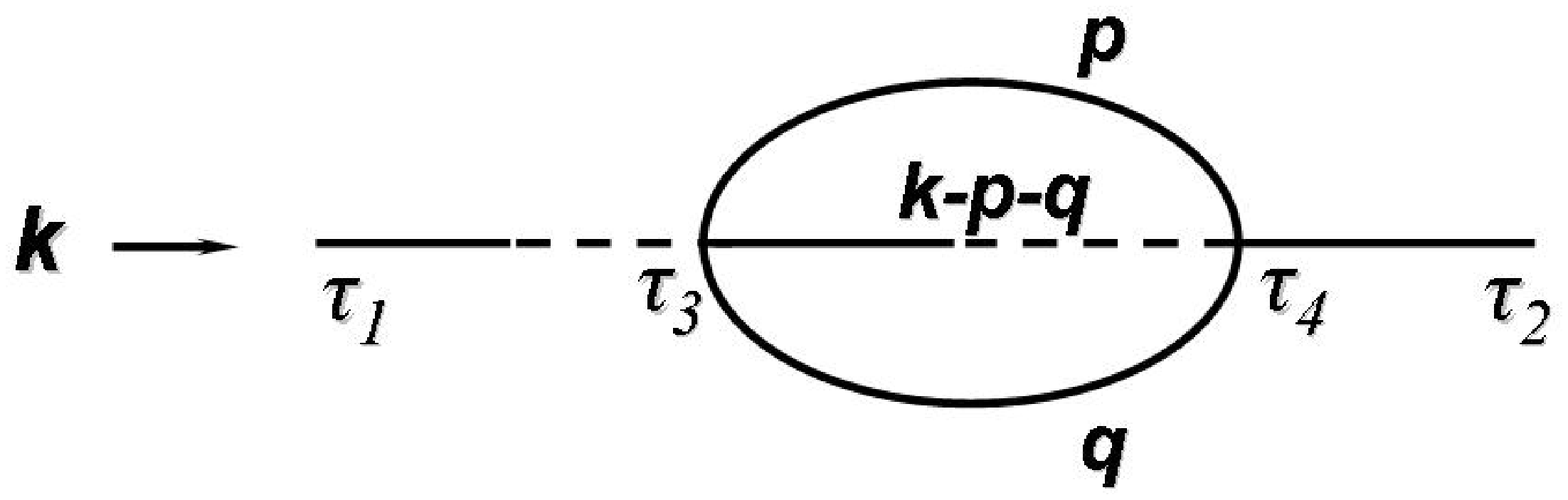}  \includegraphics[width=.3\textwidth]{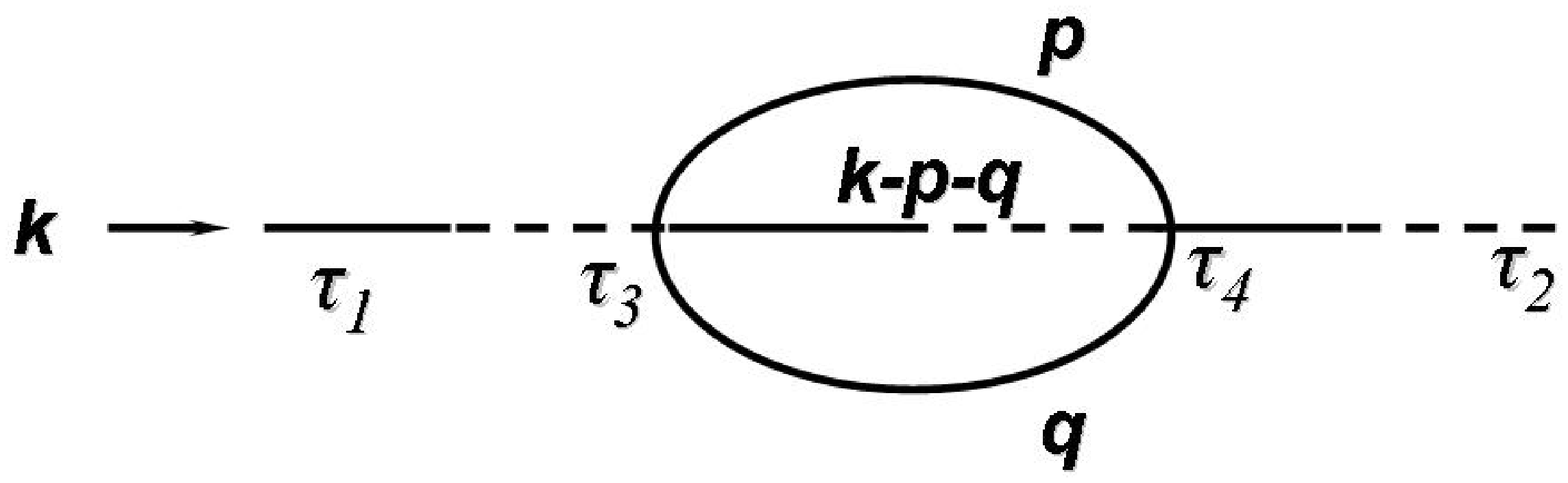} \\
\caption{Second order sunrise diagrams  for $G_C$({\it left}) and $G_R$({\it right}).}\label{sunrise2l}
\end{figure}

\noindent
The two-tadpole chain graphs can easily be calculated, basically adding a $G_R$ and closing in a
tadpole two of the straight lines of the second vertex (fig. \ref{G2l}).
The amplitudes for the two-tadpole chain propagators are
\begin{eqnarray}\label{gc2}
\mathrm{G}_C^{(2)}(k,\tau_1,\tau_2)&\simeq & \frac{H}{2 k^3} \left(\frac{\lambda \Lambda}{ H^2}\right)^2 \frac{1}{2!} (\ln^2(-k\tau_1) + \ln^2(-k\tau_2)),
\end{eqnarray}
\begin{eqnarray}\label{gr2}
-i\mathrm{G}_R^{(2)}(k, \tau_1,\tau_2) & \simeq& \frac{-iH^2}{3} \left(\frac{\lambda \Lambda}{H^2}\right)^2 \frac{1}{2!}({\tau_1}^3 - {\tau_2}^3)\ln^2\left(\frac{\tau_1}{\tau_2}\right) \, .
\end{eqnarray}	
The $n$-th order chain graphs can be calculated in the same way and we find
\begin{eqnarray}\label{Gcchain}
\mathrm{G}_C^{\rm chain}(k, \tau_1,\tau_2) & \equiv & \frac{H^2}{2 k^3} e^{\frac{\lambda \Lambda}{ H^2}\ln(k^2 \tau_1 \tau_2)}\nonumber \\
& = & \mathrm{G}_C^{(0)} e^{\epsilon \ln(k^2 \tau_1 \tau_2)}\\
&=&\mathrm{G}_C^{(0)} (k^2 \tau_1 \tau_2)^\epsilon \, ,
\end{eqnarray}
\begin{equation}\label{Grchain}-i\mathrm{G}_R^{\rm chain}(k,\tau_1,\tau_2)  \equiv  \theta(\tau_1 - \tau_2) \frac{-iH^2}{3} ({\tau_1}^{3-\epsilon}{\tau_2}^{\epsilon} - {\tau_1}^{\epsilon}{\tau_2}^{3 - \epsilon}),
\end{equation}
where $\epsilon  = \frac{\lambda \Lambda}{ H^2}$. The IR modes  resummation of the  chain diagrams generates
a spectrum of perturbations which is no longer flat, but blue tilted. Next, we consider the resummation of the
tower graphs.

\subsubsection{Tower Graphs}
\noindent
To write the amplitude for the $\mathrm{G}_C^{\rm tower}$ at second order we must be more careful. Indeed, we have two loops which are chained one into the other; in the $G_R$ tower diagram from two consecutive retarded propagators $\mathrm{G}_R^{(0)}(k,\tau_1,\tau_3)\mathrm{G}_R^{(0)}(k,\tau_3,\tau_2)$ one obtains $\tau_1>\tau_3>\tau_2$ while for the $G_C$ tower diagram one obtains only $\tau_1>\tau_3$. We note however that the times internal to the loops do not receive constraints from the $\theta$ functions relative to $\tau_1,\tau_3,\tau_2$. The integral over the time $d\tau_4$ must then extend from a loop characteristic time to the upper end which is given by the $\theta$ function embedded in $G_R(p, \tau_3, \tau_4)$. The only time scale available is the one given by the momentum $p$, thus the integral over $d\tau_4$ is evaluated between $-\frac1p$ and $\tau_3$. With these considerations the amplitude for the tower $G_C$ diagram is given by
\begin{eqnarray}
\mathrm{G}_{C}^{tower(2)}(k,\tau_1,\tau_2) & = & \int_{-\frac{1}{k}}^{\tau_1} d\tau_3 \int \frac{d^3p}{(2\pi)^3} \int\frac{d^3q}{(2\pi)^3} \int_{-\frac{1}{p}}^{\tau_3}d\tau_4 (-i)\mathrm{G}_R^{(0)}(k,\tau_1,\tau_3) \nonumber \\
& &\left(-i\lambda a^4(\tau_3)\right)  \mathrm{G}_C^{(0)}(k,\tau_3,\tau_2) (-i)\mathrm{G}_R^{(0)}(p,\tau_3,\tau_4)\nonumber \\
& & \left(-i\lambda a^4(\tau_4)\right) \mathrm{G}_C^{(0)}(q,\tau_4,\tau_4) \mathrm{G}_C^{(0)}(p,\tau_4,\tau_3) \, ,
\end{eqnarray}
and the amplitude for the retarded propagator is
\begin{eqnarray}
\mathrm{G}_{R}^{tower(2)}(k,\tau_1,\tau_2) & = & \theta(\tau_1-\tau_2)\int_{\tau_2}^{\tau_1} d\tau_3 \int \frac{d^3p}{(2\pi)^3} \int\frac{d^3q}{(2\pi)^3} \int_{-\frac{1}{p}}^{\tau_3}d\tau_4 (-i)\mathrm{G}_R^{(0)}(k,\tau_1,\tau_3) \nonumber \\
& &\left(-i \lambda a^4(\tau_3)\right)  \mathrm{G}_R^{(0)}(k,\tau_3,\tau_2) (-i)\mathrm{G}_R^{(0)}(p,\tau_3,\tau_4)\nonumber \\
& & \left(-i\lambda a^4(\tau_4)\right) \mathrm{G}_C^{(0)}(q,\tau_4,\tau_4) \mathrm{G}_C^{(0)}(p,\tau_4,\tau_3) \, ,
\end{eqnarray}
Performing the calculation and considering the mirror graph we obtain
\begin{equation}
\mathrm{G}_C^{{\rm tower} (2)}  \simeq  \frac{H^2}{2 k^3} \left(\frac{\lambda \Lambda}{H^2}\right)^2 \frac{1}{2!} [\ln^2(-k\tau_1) + \ln^2(-k\tau_2)]\, ,
\end{equation}
while for the retarded propagator we obtain:
\begin{equation}\label{gr2tower}
-i \mathrm{G}_R^{{\rm tower}(2)} \simeq \theta(\tau_1 - \tau_2) \frac{ i H^2}{3} \left(\frac{\lambda \Lambda}{ H^2}\right)^2  \left({\tau_1}^3+{\tau_2}^3\right) \ln^2 \left(\frac{\tau_1}{\tau_2}\right)\, .
\end{equation}
Also in this case it is possible to calculate the $n$-th order contribution:
\begin{displaymath}
-i\mathrm{G}_R^{{\rm tower} (n)}(k, \tau_1, \tau_2) = \theta(\tau_1 - \tau_2) \frac{iH^2}{3} \epsilon^{2+m} \frac{2^{2m}}{(2+m)!}
 ({\tau_1}^3 + {\tau_2}^3) \ln^{2+m} \left( \frac{\tau_1}{\tau_2}\right) \, ,
\end{displaymath}
and summing over all contribution with $m$ between 0 and $\infty$ one obtains a contribution proportional to
the one-loop tadpole diagram. For example, for the retarded propagator:
\begin{equation}
-i\mathrm{G}_R^{\rm tower}(k, \tau_1, \tau_2)=
\theta(\tau_1-\tau_2)\frac{iH^2}{3} ({\tau_1}^3 + {\tau_2}^3) 4\epsilon \ln \left( \frac{\tau_1}{\tau_2}\right)\, .
\end{equation}
Performing this last sums is equivalent to sum {\it vertically} over the whole class of tower graphs. Interestingly,
the result is proportional to the first order graph, see eq. (28), that can be resummed as shown in the previous section.
Therefore the resummation of the tower graphs can be accounted for  just  sending $\epsilon$ into $\epsilon' = 5\epsilon$,
which does not change the properties obtained from  eq. (\ref{Gcchain}).

\subsubsection{Sunrise Graphs}
\noindent
Despite the graphical difference in respect to the tadpole chains, these diagrams translate exactly into  tadpole graphs. For example the sunrise $G_C$ can be written as
\begin{eqnarray}\label{G2sun}
\mathrm{G}_{C}^{{\rm sun}(2)}(k,\tau_1,\tau_2) & = & \int_{-\frac{1}{k}}^{\tau_1} d\tau_3 \int \frac{d^3p}{(2\pi)^3}\int \frac{d^3q}{(2\pi)^3} \int_{-\frac{1}{k}}^{\tau_3}d\tau_4 (-i)\mathrm{G}_R^{(0)}(k,\tau_1,\tau_3) \nonumber \\
& &\left(-i \lambda a^4(\tau_3)\right)  \mathrm{G}_C^{(0)}(p,\tau_3,\tau_4) (-i)\mathrm{G}_R^{(0)}(k-p-q,\tau_3,\tau_4)\nonumber \\
& & \left(-i \lambda a^4(\tau_4)\right) \mathrm{G}_C^{(0)}(q,\tau_3,\tau_4) \mathrm{G}_C^{(0)}(k,\tau_4,\tau_2)\, , \\
\nonumber
\end{eqnarray}
and the same for the $G_R$.  However the the combinatorial coefficient in front of the integrals is different than in the case of chain graphs, since there are two ways to contract $\phi_C(\tau_1)$ with a $\phi_\Delta$ of the vertices, then three to contract one $\phi_C$ of the first vertex to the $\phi_\Delta$ of the second and finally six to contract the remaining free $\phi_C$  in the two vertices. In total we have a $36/2!$.
The contributions coming from the tadpole and sunrise diagrams  differ for a numerical constant. If we
 resum the whole class of sunrise diagrams the result is therefore proportional to the first order tadpole graph.
One should not claim victory too soon though.
Already at one loop, one should account for the vertex renormalization. Unfortunately,
the IR resummation of the vertex renormalizing graphs
 proves to be a difficult task because of the presence of diagrams whose time flow
in the internal lines is not continous. This is not surprising at all, since it is very well known that only
in certain class of self-interacting models, the full resummation is possible. We now turn to one of these
examples, generalizing our results to $N\gg 1$ fields respecting an $O(N)$  symmetry.

\section{$O(N)$ Symmetry and the trispectrum}
\begin{figure}[!h]
\centering
\includegraphics[width=.15\textwidth]{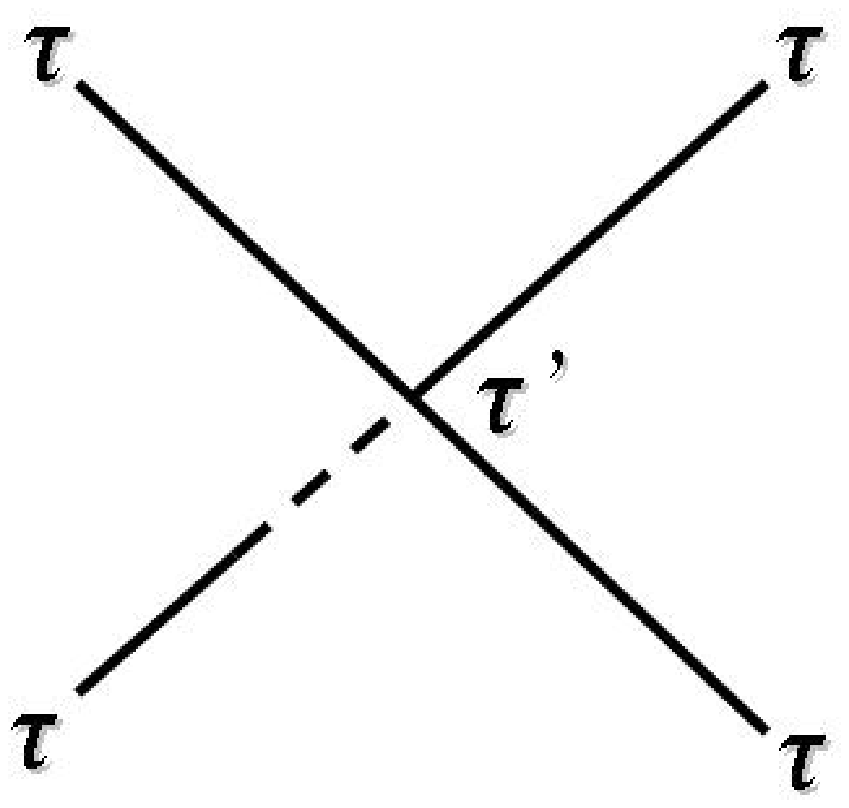}  \quad  \includegraphics[width=.17\textwidth]{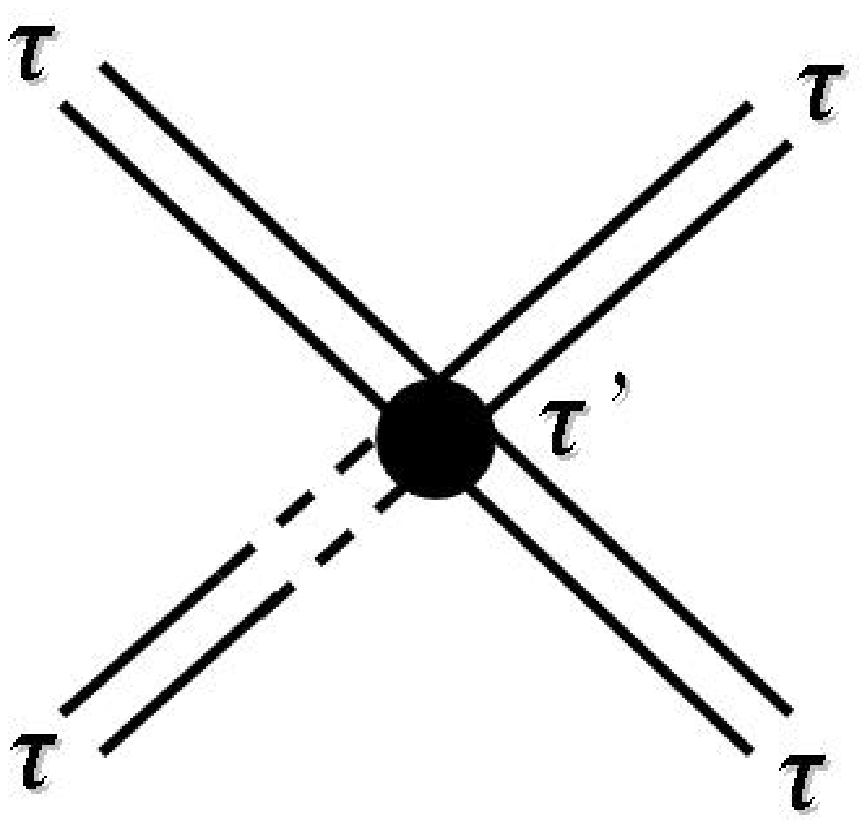}
\caption{Four point function calculated at tree level ({\it left})  and with the resummed propagators ({\it right})  .}\label{4tree}
\end{figure}

One way to be able to discard all the graphs except towers and tadpoles is to assume that we have $N\gg 1 $
fields with an $O(N)$ symmetry. Under this assumption, thanks to the normalization of the vertices (that gives a $N^{-1}$ for each vertex) and the loop over the free indices in loops (an $N$ for every free index) , tadpole and tower graphs are proportional to $N^0$, while all the other graphs scale at least as $N^{-1}$. Therefore, under the assumption that $N\gg 1$ and
according to our diagrammatic results, we conclude that IR effects may be resummed.
In Ref. \cite{slothriotto} the same model was analyzed  using non perturbative stochastic techniques. There, from the
Fokker-Planck equation, the field quadratic mean value, which is linked to the correlation functions by   $\langle\phi^2\rangle=G^{++}(x,x)$,was obtained
\begin{eqnarray} \label{sol1}
\left< \left(\frac{\phi}{H}\right)^{2}\right> = G^{++}(x,x) = \frac{\textrm{Tanh}(\frac{\sqrt{\bar\lambda}}{4\pi^2}\ln a)}{\sqrt{\bar\lambda}} \, ,
\end{eqnarray}
with $\bar \lambda = 4\pi^2\lambda/3$, and then inserted it into the gap equation in order to be able to solve for  $G^{++}(x,x')$,
\begin{eqnarray} \label{gln}
-\left(\square_x + m^2 + \frac{\lambda}{2}
\left[ \phi_c^2(x) + G^{++}(x,x) \right]  \right)
G^{++}(x,x') =   i
\frac{\delta(x-x')}{\sqrt{-g}}.
\end{eqnarray}
The result for the two-point correlation functions was
\begin{eqnarray}
G_C(k;\tau_1,\tau_2) 
\approx \frac{H^2}{2k^3}(-k\tau_1)^{\delta}(-k\tau_2)^{\delta},
\end{eqnarray}
\begin{eqnarray}
G_R(k;\tau_1,\tau_2) \approx \theta(\tau_1-\tau_2)\frac{H^2}{k^3}\left[(-k\tau_1)^{\delta}(-k\tau_2)^{3-\delta}-(-k\tau_2)^{\delta}(-k\tau_1)^{3-\delta})\right]~.
\end{eqnarray}
which have the same form as our eq. (\ref{Gcchain}) and (\ref{Grchain}), except for the $\delta \equiv m_{\rm np}^2/3H^2$. The mass $m_{\rm np}$ is what controls the IR divergences and appears because $G^{++}(x,x)$ goes rapidly to a constant and thus plays the role of a mass in eq.(\ref{sol1}).  In our diagrammatic approach, the same non perturbative mass appears
if we think to eq. (22) as a gap equation for the IR cut-off, but using the resummed propagator $G_C^{\rm chain}$,

\begin{eqnarray}
m_{\rm np}^2 = 3 \lambda  \Lambda = \frac{9\lambda H^4}{8\pi^2 m_{\rm np}^2},
\end{eqnarray}
that is
\be
 m_{\rm np}^2 =  \frac{3H^2}{2\pi} \sqrt{\frac{\lambda}{2}},\,\,\epsilon = \frac{1}{2\pi}\sqrt{\frac\lambda2},
 \ee
which has the same dependence on $\lambda$ as the non perturbative mass found in \cite{slothriotto}. The numerical coefficient do not coincide due to the different normalization of the potential. Indeed, redefining $\lambda \to \lambda/3!$, one obtains the potential $V= \lambda\phi^4 /4$ and $m_p$ and $\epsilon$ coincide with $m_{\rm np}$ and $\delta$ of Ref. \cite{slothriotto}.
Therefore, the diagrammatic approach reproduces
the findings obtained using  the stochastic approach. In particular, if we are interested in the evaluation of the
trispectrum, we need to evaluate

\be
\langle \delta\phi_C(\bkone, \tau)\delta\phi_C(\bktwo, \tau)\delta\phi_C(\bkthree, \tau)\delta\phi_C(\bkfour, \tau)\rangle  =  T(\bkone,\bktwo,\bkthree,\bkfour) \picube \sdelta{\bkone+\bktwo+\bkthree+\bkfour}.
\end{equation}
At tree level the 4-point function is built with free propagators as shown in figure (\ref{4tree})
It is necessary to sum over the permutations, since each of the four momenta can be flowing through
the retarded propagator. The amplitude at tree level is
\begin{eqnarray}
 T^{tree}(\bkone,\bktwo,\bkthree,\bkfour)& = & \sum_{i=1}^4 \int^{\tau} d\tau' e^{-i(\sum_{l=1}^4 k_l)\tau'}
\left(-i \frac{\lambda}{N} a^4(\tau')\right) \\
& &(-i\mathrm{G}_R^{(0)}(k_i, \tau, \tau'))\prod_{j \neq i} \mathrm{G}_C^{(0)}(k_j, \tau, \tau')
e^{-i[\sum_{i=1}^4 k_i] \tau'},   \nonumber
\end{eqnarray}
which, for $k\tau_i\ll 1$ ($i=1,\cdots, 4$):
\begin{equation}\label{tree}
 T^{\rm tree}(\bkone,\bktwo,\bkthree,\bkfour, \tau) =  -\frac{\lambda H^4}{24 N \prod_{i=1}^{4}{k_i}^3}
\sum_{i=1}^{4} {k_i}^3 \left[-\gamma -\frac{i \pi }{2}  -\ln\left[-(\sum k_i)\tau\right] \right]\, .
\end{equation}
Equation (\ref{tree}) reproduces the result obtained by Bernardeau {\it et al.} in \cite{ber}.
Under the hypothesis of $N\gg 1$,  the trispectrum  is obtained the exactly resummed propagators,  that is
the 'double-line' propagators  on the right of fig.~\ref{4tree}.
The amplitude is
\begin{eqnarray}
 T^{\rm chain}(\bkone,\bktwo,\bkthree,\bkfour)& = & \sum_{i=1}^4 \int^{\tau} d\tau'
e^{-i(\sum_{l=1}^4 k_l)\tau'} \left(-i \frac{\lambda}{N} a^4(\tau')\right) \\
& &(-i\mathrm{G}_R^{\rm chain}(k_i, \tau, \tau'))\prod_{j \neq i} \mathrm{G}_C^{\rm chain}(k_j, \tau, \tau') e^{-i[\sum_{i=1}^4 k_i] \tau' }\nonumber \\
& = & -\frac{\lambda}{N}\frac{H^4}{24} \frac{1}{\prod_{i=1}^4 {k_i}^3} \sum_{i=1}^4 {k_i}^3 \nonumber\\
& & \prod_{j\neq i} {k_j}^{2\epsilon}\tau^{6\epsilon} \left[-E_{4-4\epsilon}( i k_t \tau) + E_{1-2\epsilon}(i k_t \tau) \right]  \, ,
\end{eqnarray}
where the function ${E_n}(z)$ is defined as
\be
{E_n}(z) = \int_1^\infty \frac{e^{-zt}}{t^n} dt.
\ee
Thus in the limit $-k\tau_i\ll 1$ the amplitude becomes
\be
\label{final4}
 T^{\rm chain}(\bkone,\bktwo,\bkthree,\bkfour) =  \frac{\lambda}{N}\frac{H^4}{48\epsilon}
\frac{\sum_{i=1}^{4}k_i^3}{\prod_{i=1}^{4}k_i^3},
\ee
which again coincides in form with what found in Ref. \cite{slothriotto}. In particular, the resummed trispectrum shows an enhancement
factor $\sim 1/\sqrt{\lambda}$ compared to the tree level result. Higher loop corrections to the trispectrum are suppresed
by the fact that the propagators are now IR regulated.

\section{Conclusions}\label{conclusions}
\noindent
In this paper we have analyzed the IR corrections to the correlators for a self-interacting scalar field
in a de Sitter background. We have used a full diagrammatic approach within the in-in formalism. In this sense, our results
should be considered complementary to alternative approaches, {\it e.g.} the stochatic approach. It is reassuring
that the same results are obtained once the IR resummation can be performed as in the large $N$ $O(N)$ theory.
\section{Acknowledgments}
The author would like to thank Prof. A. Riotto for the invaluable guidance and helpful and enlightening discussions during the preparation of this work.

\end{document}